\documentclass[%
 amsmath,amssymb,
 aps,
 prx,twocolumn,english,superscriptaddress,floatfix,longbibliography
]{revtex4-2}

\usepackage{graphicx}% Include figure files
\usepackage{dcolumn}% Align table columns on decimal point
\usepackage{bm}% bold math
\usepackage{enumitem}
\usepackage[caption=false]{subfig}
\usepackage{dsfont}
\usepackage{accents}
\usepackage{hyperref}
\usepackage{cleveref}

\usepackage{tikz}
\newenvironment{captivy}[1]{%SETUP
  \begin{tikzpicture}[every node/.style={inner sep=0}]
    \node[anchor=south west,inner sep=0] (image) at (0,0) {#1};
    \begin{scope}[x={(image.south east)},y={(image.north west)}]
}%
{%TEARDOWN
        \end{scope}%
  \pgfresetboundingbox
  \path[use as bounding box] (image.south west) rectangle (image.north east);
  \end{tikzpicture}%
}

\newcommand*{\oversubcaption}[3]{%
  \draw (#1) node[fill=white,inner sep=0pt, opacity=0.2, above, yscale=1.1, xscale=1.1] {\phantom{(a)#2}};
  \draw (#1) node[inner sep=0pt, above]{%
    \subfloat[#2\label{#3}]{\phantom{(a)}}
  };
}

\Crefname{figure}{Fig.}{figures}
\Crefname{equation}{Eq.}{equations}
\creflabelformat{equation}{#2\textup{#1}#3}

\begin{document}

\preprint{APS/123-QED}

\title{Exploring canyons in glassy energy landscapes using metadynamics}% Force line breaks with \\
%\thanks{A footnote to the article title}%

\author{Amruthesh Thirumalaiswamy}
%  \altaffiliation[Also at ]{Department of Chemical and Biomolecular Engineering, 
%  Univeristy of Pennsylvania.}%Lines break automatically or can be forced with \\
\author{Robert A. Riggleman}
 \email{rrig@seas.upenn.edu}
%  \altaffiliation[Also at ]{Department of Chemical and Biomolecular Engineering, 
%  Univeristy of Pennsylvania.}%Lines break automatically or can be forced with \\
\author{John C. Crocker}%
 \email{jcrocker@seas.upenn.edu}
\affiliation{%
 Department of Chemical and Biomolecular Engineering, University of Pennsylvania, Philadelphia, Pennsylvania\\
%  This line break forced with \textbackslash\textbackslash
}%

%\date{\today}% It is always \today, today,
             %  but any date may be explicitly specified

\begin{abstract}
The complex physics of glass forming systems is controlled by the structure of the low energy portions of their potential energy landscapes. Here, we report that a modified metadynamics algorithm efficiently explores and samples low energy regions of such high-dimensional landscapes. In the energy landscape for a model foam, our algorithm finds and descends meandering `canyons' in the landscape, which contain dense clusters of energy minima along their floors. Similar canyon structures in the energy landscapes of two model glass formers---hard sphere fluids and the Kob-Andersen glass---allow us to reach high densities and low energies, respectively. In the hard sphere system, fluid configurations are found to form continuous regions that cover the canyon floors up to densities well above the jamming transition. For the Kob-Andersen glass former, our technique samples low energy states with modest computational effort, with the lowest energies found approaching the predicted Kauzmann limit.

% The column width is: \the\columnwidth 510.0pt
\end{abstract}

\maketitle

% The column width is: \the\columnwidth: 246 p.t.

%Intro
\section{\label{sec:Introduction}Introduction}
Many outstanding problems in understanding glasses have been related to their potential energy landscapes \cite{PELWales2007}, hypersurfaces describing a system’s total potential energy spanning the high-dimensional configuration space formed by all the particles’ spatial coordinates \cite{LiqStrucStillinger1982}. These landscapes have a complex geometry, as indicated by a broad distribution of energy basin hypervolumes \cite{PESMassen2007, BasinsXu2011}, highly tortuous steepest descent paths \cite{BasinAshwin2012, SGMHwang2016}, and fractal clustering of local minima \cite{SGMHwang2016}. Rapid quenching from a very high temperature (equivalent to following a steepest descent path from a random configuration) reports the landscape's local minima, or inherent structures (ISes) \cite{LiqStrucStillinger1982}, each weighted by their associated drainage basin hypervolume \cite{PESMassen2007, BasinsXu2011}. The ensemble of such quenched configurations \cite{JammingLiu2010} having the largest basin hypervolumes are distinctly different from the lowest energy inherent structures \cite{HSBerthier2016, HSOzawa2017} that control glass transitions \cite{SimulatedAnnealinSastry1998, FractalCharbonneau2014, IdealGlassRoyall2018}. Navigating such high-dimensional spaces and mapping the arrangement of glassy states remains a major challenge. Further, the heterogeneous nature of glassy dynamics \cite{dyheteroBerthier2011} makes the use of collective descriptors of system dynamics ineffective. Methods like eigenvector-following \cite{TSWilliam1981}, and techniques for exhaustively enumerating inherent structures \cite{Wales2018} allow the spatial arrangement of glassy states to be explored, but tend to be computationally expensive \cite{BHWales1999, CrystalKAWales2001} in large systems. Swap Monte Carlo \cite{HSBerthier2016, HSOzawa2017, KAParmar2020} and similar methods allow the canonical sampling of glassy states, but jump around configuration space, obscuring the states' arrangement, and these methods are ineffective in bonded systems. Meanwhile, optimizers such as basin-hopping \cite{BHWales1997} efficiently find the lowest states but operate on a modified landscape without barriers, overlooking interesting characteristics of the landscape in the process.
% The column width is: \the\columnwidth: 246.09 p.t.

Here, we modify a high-dimensional metadynamics \cite{Laio2002} algorithm by Yip \cite{MetadynamicsKushima2009} and use it to discern the arrangement of inherent structures in glassy energy landscapes; calling our approach the Metadynamics-Inspired Multifractal Sampling Explorer (MIMSE). Applying this algorithm to a model of foams or soft glassy materials (SGMs) \cite{SGMHwang2016}, it finds meandering, smooth-walled canyon-like structures in the landscape, and descends into them to find many inherent structures clustered on the canyons' floors. Analyzing small ensembles of biased trajectories reveals that these canyons have well-defined widths, and become narrower as they meander to progressively lower energies. To apply this approach to a popular model of dense hard-sphere (HS) fluids \cite{HSBerthier2016}, we examine the corresponding extended energy landscape for compressible or soft spheres. The algorithm again finds and descends similar canyons, finding HS configurations (where the soft sphere potential energy is zero) that cover the canyon floor. These zero energy configurations form a continuous, connected domain, that can be effectively found up to volume fractions as high as $0.68$. However, locating these HS states from random points in configuration space becomes exponentially costly at higher densities. Last, when applied to a simple model of an atomic glass former, the Kob-Andersen (KA) model \cite{KA1995}, our algorithm again finds and rapidly descends canyons, reaching very low energies with modest computational effort. Exploring many canyons yields a distribution of canyon bottom energies the lowest of which approach the predicted Kauzmann energy limit \cite{KauzmannAngell1999}. Surprisingly, the low-dimensional `canyon floors' that contain the glassy configurations were found to have the same effective fractal dimension in all three systems' landscapes. 

\section{\label{sec:MIMSE}The MIMSE Algorithm}

MIMSE is an athermal, metadynamics based approach \cite{Laio2002} directly applied to the $3N$-dimensional potential energy landscape of $N$ particles, and uses sequentially added bias potentials to overcome energy barriers \cite{MetadynamicsKushima2009}. We use the Fast Inertial Relaxation Engine (FIRE) \cite{StrucRelaxBitzek2006} to relax the system configuration on the biased energy landscape; tests using steepest descent gave similar results but at higher computational cost. The bias potential employed is a $3N$-dimensional, bell-shaped, smooth quartic function with the form
\begin{equation}
  U(\mathbf{r})=\begin{cases}
     \mathcal{U}_0{\left(1 - {\left(\frac{\lVert{\underaccent{\bar}{\mathbf{r}} - \underaccent{\bar}{\mathbf{r}}_\text{m}}\rVert}{\mathcal{U}_\sigma}\right)}^2\right)}^2, & \text{if $\lVert{\underaccent{\bar}{\mathbf{r}} - \underaccent{\bar}{\mathbf{r}}_\text{m}}\rVert<\mathcal{U}_\sigma$}\\
    0, & \text{otherwise}.
  \end{cases}
  \label{eq:bias potential}
\end{equation}
Here, $\underaccent{\bar}{\mathbf{r}}$ represents the $3N$ dimensional system configuration, and $\underaccent{\bar}{\mathbf{r}}_\text{m}$ the IS minimum location around which the bias is centered. The bias parameters $\mathcal{U}_\sigma$ and $\mathcal{U}_0$, represent the $3N$-dimensional radial extent and energetic height of the bias respectively. To ensure that the center of mass of the system remains at rest, we modify the corresponding bias forces to ensure the total force on the system is zero. To ensure stable configuration dynamics, we use models with energy landscapes that are continuous and differentiable.

Algorithmic flow starts at a local minimum of the landscape, sampled typically from the quenched ensemble, described earlier. Starting at such a minimum, a bias is added centered on its location, forming a local maximum on the biased landscape. To move away from this unstable maximum, the system is given a small, random $3N$-dimensional displacement (corrected so as to preserve the position of the system's center of mass). This configuration is then relaxed to the nearest minimum on the biased landscape using FIRE, and the process repeated by adding a new bias at each subsequent minimum. 

In conventional metadynamics, the bias radius is smaller than the separation between neighboring ISes, leading to basin filling and exhaustive enumeration of minima \cite{Laio2002, MetadynamicsKushima2009}. Here, we use larger bias radii, such that when a bias is added, many ISes within $\mathcal{U}_\sigma$ of $\underaccent{\bar}{\mathbf{r}}_\text{m}$ are covered by that bias, effectively forcing the system over their associated nearby energy barriers. Values of $\mathcal{U}_\sigma$ and $\mathcal{U}_0$ that lead to efficient landscape descent in energy are found by manual `tuning' using a divide and conquer-like approach, see \hyperref[sec:Methods]{Methods}. For computational efficiency, a $3N$-dimensional neighbor list is maintained to track the different biases affecting the system at its current $3N$-dimensional position. In practice, we find that reasonably `old' biases can often be retired without difficulty, keeping computational and memory costs manageable. Additional details of the algorithm are provided in \hyperref[sec:Methods]{Methods}.

As the algorithm proceeds, we distinguish among the minima encountered and store and analyze the subset which lie outside the bias potentials, which also correspond to physical inherent structures (ISes) of the unbiased energy landscape. As we will primarily report below, this approach consistently yields ISes in glassy landscapes.

\section{\label{sec:SGM}SGM Energy Landscape}

\begin{figure}[t]
  \centering
  \begin{captivy}{\includegraphics[scale=1]{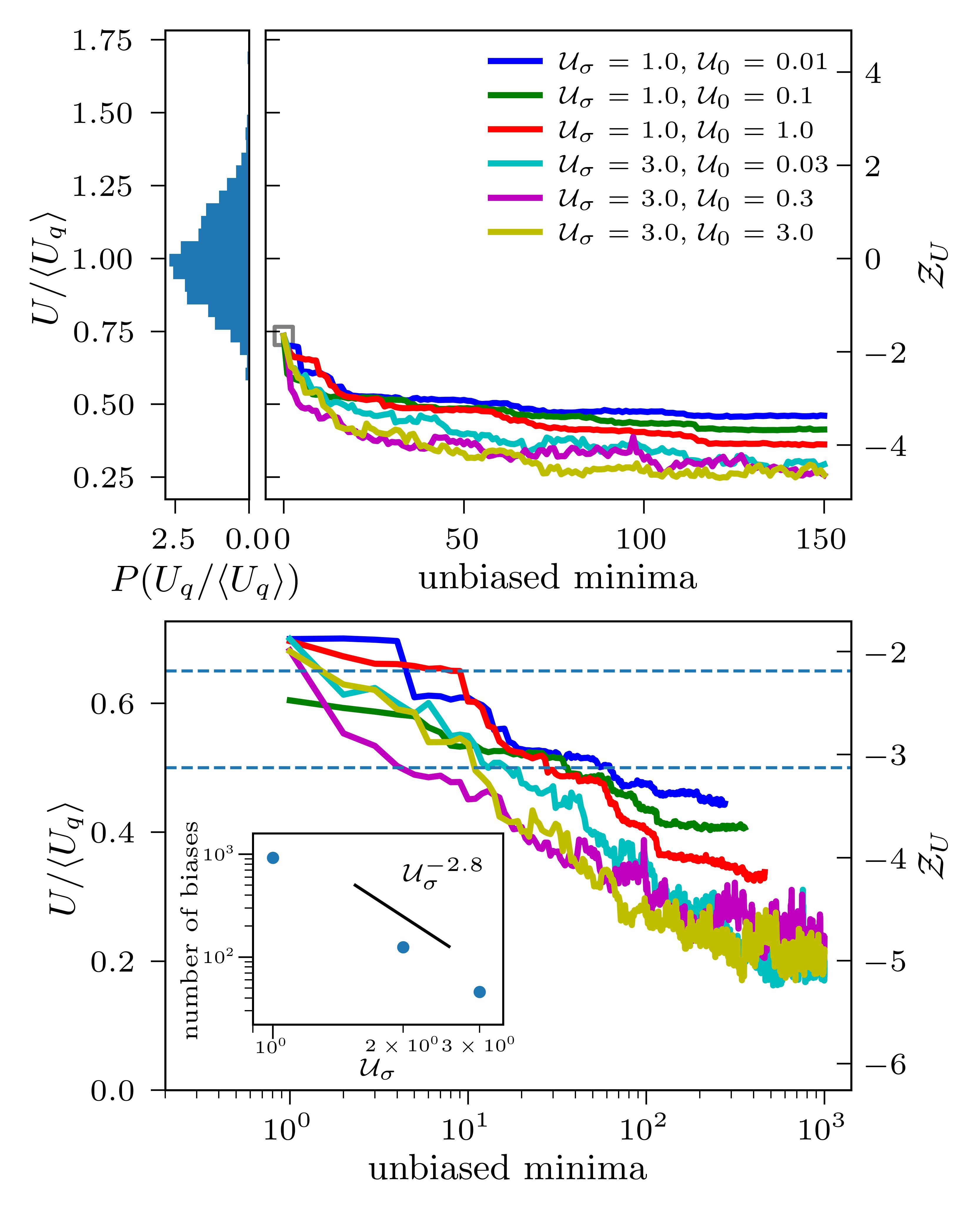}}
    \oversubcaption{0.05, 0.98}{}{fig:SGMa}
    \oversubcaption{0.05, 0.48}{}{fig:SGMb}
  \end{captivy}
  \caption{Exploring a soft-glassy material landscape with MIMSE finds low-energy inherent structures efficiently.
  (a) The left panel shows the probability distribution of energies for an ensemble of $1000$ different `quenched' initial configurations, formed by FIRE relaxation of random configurations. Applying MIMSE, the energy of the configurations decreases rapidly, depending on bias radius, $\mathcal{U}_{\sigma}$. The right axis rescales the energy relative to the mean and standard deviation of the quenched energies $U_q$, also known as the \textit{z}-score, $\mathcal{Z}_{U}$.
  (b) The descending energy trajectory appears roughly logarithmic in time.
  (Inset) The number of biases that are required to descend a given energy range (dashed lines in \Cref{fig:SGMb}) shows a strong dependence on $\mathcal{U}_{\sigma}$, which can be approximated by a power-law over a narrow range.}
  \label{fig:SGM}
\end{figure}

We first demonstrate our approach on a model for foam or soft glassy materials (SGM) \cite{SGMSollich1997, SGMSollich1998, SGMSollich2012}, whose energy landscape was characterized in an earlier study \cite{SGMHwang2016}. The system consists of a highly polydisperse collection of $N \sim 350$ compressible `soft' spheres interacting via a purely repulsive harmonic interactions when they overlap. We choose a volume fraction, $\phi = 0.75$, about $0.03$ higher than the threshold for solidity or `jamming', $\phi_J^{SGM}$, in this model \cite{SGMHwang2016}. The distribution of radii ($R$) resembles that of bubbles in a ripening foam, described by a Weibull distribution with polydispersity $\Delta = ( \langle R^2 \rangle - {\langle R \rangle}^2)^{1/2}/\langle R \rangle \simeq 0.64$, see \hyperref[subsecSI:SGM]{\textit{SI Appendix}} for model details. To create initial configurations of this model, we select random points in configuration space and relax them to their nearest energy minimum using FIRE. This ensemble of `quenched' inherent structures has a roughly Gaussian distribution of energies, see \Cref{fig:SGMa} (left panel).

Our key finding is that for certain bias parameters, MIMSE rapidly descends down the potential energy landscapes of glassy systems. Typical results for the SGM model landscape are shown in \Cref{fig:SGM}, reaching energies that are much lower than the energies of quenched configurations after only a few hundred biases. These proceed to lower energy logarithmically with additional effort, reaching energies which, compared to the quenched energy distribution, are $\mathcal{Z}_U \approx -5$ standard deviations below the mean of the initial quenched inherent structures, \Cref{fig:SGM}(right scale). Notably, such significant descent in energy is only found to occur in a window of bias radii, $2 \lesssim \mathcal{U}_\sigma \lesssim 5$, and the number of biases required to descend by a given energy shows a strong dependence on the bias radius, \Cref{fig:SGMb}(inset). Larger biases descend the landscape far faster than smaller ones, up to a maximum radius. The energetic height of the bias $\mathcal{U}_0$ appears less critical, descent down the energy landscape is observed in a wide window, $10^{-3} \lesssim \mathcal{U}_0/\mathcal{U}_\sigma \lesssim 10^3$.

Our first task is to understand what features of the SGM landscape enable our algorithm to descend to such low energies, and to explain the narrow window of bias radii where it occurs. To this end, we formed ensembles of $1100$ FIRE trajectories, each descending the same bias around the same starting IS minimum, but initialized with different isotropically random $3N$-dimensional displacements, and examined the results as a function of bias radius. We find that, after initially moving `radially' away from the center in random directions due to the strong bias force, these paths then drift in the `angular' $(3N-1)$ dimensions according to the gradient of the underlying landscape. Finally, these paths terminate in $1100$ different energy minima near the edge of the bias.

To descend the energy landscape, the FIRE minimizer must be following gradients in the underlying landscape; perhaps the window of bias radii is due to suitable gradients only being present on corresponding lengthscales? To test this idea, we first compute the average `drift' of the ensemble of trajectories, such as would be caused by an underlying gradient. Specifically, for each trajectory we first compute a 3N-dimensional unit vector $\hat u$ pointing from the initial IS to the new energy minimum. Next, we compute the Euclidean length of the average of these unit vectors, $\|\langle\hat{u}\rangle\|$, akin to the center of mass of the new minima, as a function of bias radius, $\mathcal{U}_{\sigma}$. The results for three starting ISes at different depths in the landscape are shown in \Cref{fig:SGMclustera}. While the average drift shows a broad peak for the $\mathcal{U}_\sigma$ where energy descent is observed, a similar amount of drift is seen for much smaller and larger radii, failing to explain the lack of energy descent in those cases. 

\begin{figure}[t]
  \centering
  \begin{captivy}{\includegraphics[scale=1]{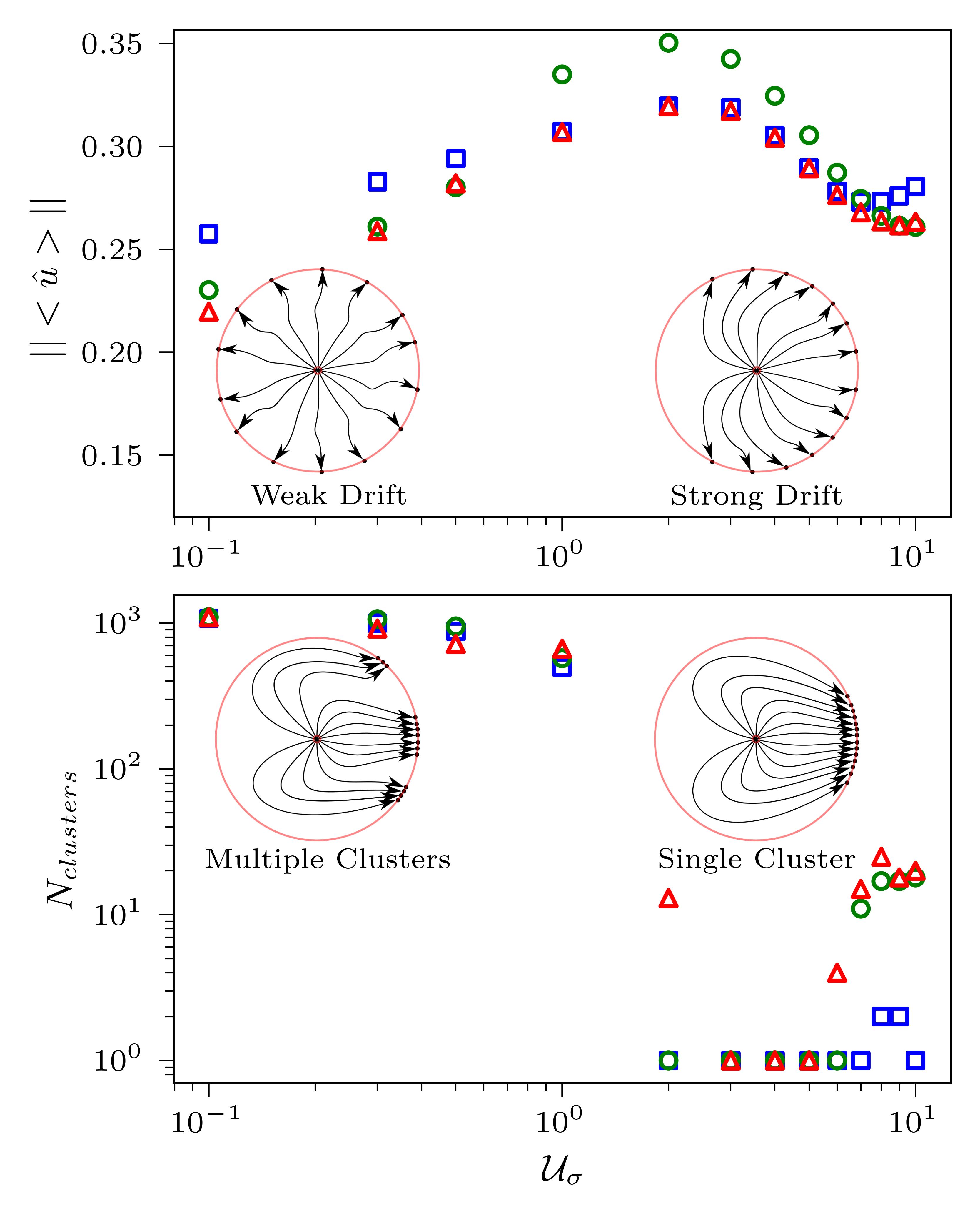}}
    \oversubcaption{0.05, 0.96}{}{fig:SGMclustera}
    \oversubcaption{0.05, 0.485}{}{fig:SGMclusterb}
  \end{captivy}
  \caption{Trajectory ensembles reveal landscape gradients and IS clustering.  Ensembles of $1100$ FIRE trajectories descending the same hyperspherical bias potential after different, isotropically random initial displacements, move radially before being affected by the underlying landscape and terminating near the bias edge (inset schematics). The final relative displacements are converted to unit vectors, $\hat{u}$.
  (a) The norm of the average of the $1100$ unit vectors $\| \langle\hat{u}\rangle \|$, reveals the average drift due to an energy gradient, for three different depths in the landscape: $\mathcal{Z}_U \approx -2.8$ (squares), $-3.8$ (circles), $-4.7$ (triangles), showing a broad peak around $\mathcal{U}_{\sigma} \approx 3$.
  (b) Performing single-linkage hierarchical cluster analysis \cite{SLCAGower1969} on the same $\hat{u}$ ensembles yields a single cluster only for $\mathcal{U}_{\sigma} \gtrsim 2$, with an upper limit that depends weakly on $\mathcal{Z}_U$. Landscape energy descent only occurs for bias radii and energies where a  single cluster is found, as explained in the text.}
  \label{fig:SGMcluster}
\end{figure}

To gather clues about descent and barrier crossing in the underlying landscape, we applied a hierarchical single-linkage clustering algorithm \cite{SLCAGower1969, OutlierSV2015} to the unit vector ensembles analyzed above. The idea is that if a subset of the trajectories are crossing a barrier, they will form a separate cluster from the others. For small and larger $U_\sigma$ values, we find multiple clusters indicating the exploration of IS clusters separated by barriers. For intermediate values of $U_\sigma$, we report that these ISes form a single extended cluster. Remarkably, we now find a one-to-one correspondence between the bias radii that give efficient energy descent and the $\hat{u}$ vectors that form a single cluster, for ISes at the three different energies considered, \Cref{fig:SGMclusterb}. Moreover, it may be noted that the maximum $\mathcal{U}_{\sigma}$ that forms a single cluster gets progressively smaller at lower energy domains in the landscape, decreasing from $7$ to $6$ to $5$ units. This observation and the correspondence between bias radius and energy descent is confirmed by another result: biases radii in the range $5-7$ rapidly descend the landscape at first, but then get `stuck', plateauing at intermediate energies, see \hyperref[subsecSI:SGM]{\textit{SI Appendix}}, \Cref{figSI:SGM}. Analysis of the same vectors using complete linkage clustering fails to resolve these clusters well, suggesting that they are extended and intertwined, rather than compact and well-separated.

These findings indicate how the algorithm works: for the appropriate bias size, it is simply following clusters of ISes like a trail of `bread-crumbs' to low energy portions of the landscape. Finding multiple clusters with too large biases indicates the crossing of large-scale barriers into other domains, losing the trail. Finding many clusters with too small biases indicates the crossing of small-scale barriers, but leaving the algorithm trapped, exploring within a dense IS cluster at nearly the same location. Lastly, a fractal scaling analysis, see \hyperref[secSI:fractalanalysis]{\textit{SI Appendix}}, of the sampled configurations suggests that the set of low energy ISes occupy a subspace with a low effective dimensionality of $\approx 2.5$.

\begin{figure}[t]
    \begin{captivy}{\includegraphics[scale=1]{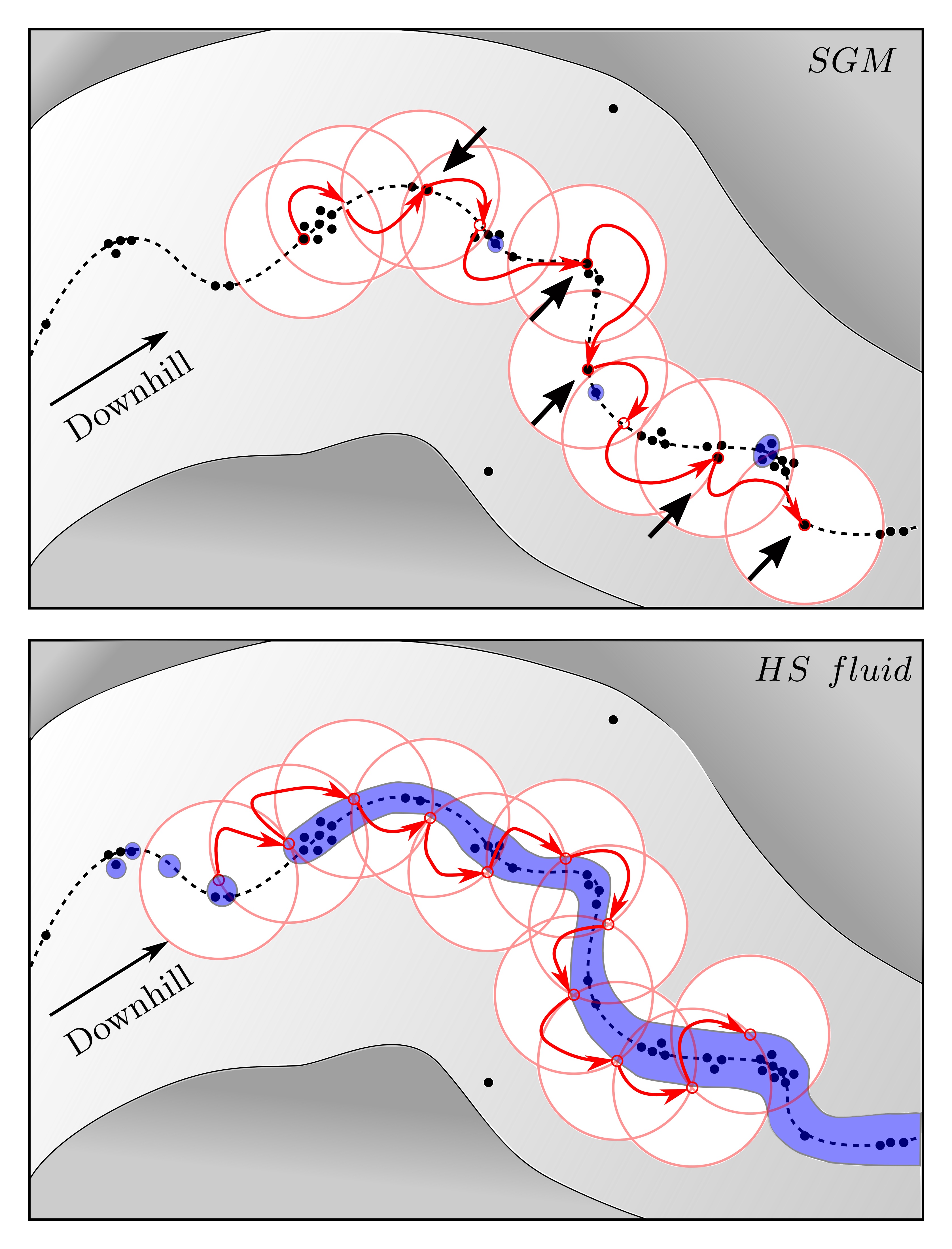}}
    \oversubcaption{0.07, 0.94}{}{fig:landscapea}
    \oversubcaption{0.07, 0.45}{}{fig:landscapeb}
  \end{captivy}
\caption{Schematic `aerial view' of the energy landscape showing `canyons' (white region) with tortuous bottoms (dashed line) descending to lower energy (left to right).
(a) For the SGM system, inherent structures (ISes) (black dots) are clustered along the canyon bottom and MIMSE skips from IS to IS (solid red path), when biases (large red circles) are sequentially applied (left to right). Red open circles represent biased minima; Unbiased minima (black arrows) are stored as output.
(b) For the HS glass, the bottom of the canyon floor is covered by a slender connected domain having zero energy (blue), as if flooded by a `stream' of water, and MIMSE skips between configurations where the bias and the zero energy domain intersect (small red circles). Some ISes in (a) may have zero energy (blue) but may be either blocked by other ISes or too rare to sample.}
\label{fig:landscape}
\end{figure}

The above analyses can be summarized schematically in \Cref{fig:landscapea}. The landscape consists of meandering canyons or tubes (as suggested by earlier studies \cite{BasinAshwin2012,SGMHwang2016}), with the ISes forming extended dense clusters along the canyon floors. These dense clusters act as traps for MIMSE when the bias radius is too small. For optimal bias radii, as shown in \Cref{fig:landscapea}, the biased FIRE path bounces off the canyon walls (finding no ISes there), evolves under the influence of the landscape gradient and ends at a new minimum further down the canyon. Sometimes this path extends out of the biased region completely, ending in a physical IS. Too large biases push the configuration into adjacent canyons, apparently preventing energy descent. Since the largest $\mathcal{U}_{\sigma}$ that descends the landscape decreases at lower energies, it appears that the canyon tapers as it descends to lower energies. Given that the canyon is high-dimensional, the reduction in its hypervolume due to even a slight decrease in width is enormous.

\section{\label{sec:HS}Hard Sphere fluid}

To determine if the energy landscapes of glass formers also contain canyon structures that resemble those in the SGM system, we apply MIMSE to two additional well-studied models. The first is a model of hard-sphere (HS) fluids \cite{HSBerthier2016, HSOzawa2017, EntropyBerthier2017}, having a size distribution given by $\textbf{P}(R) \sim R^{-3}, R \in [R_{\text{min}}, R_{\text{max}}], R_{\text{min}}/R_{\text{max}} = 0.4492$, and $N=1000$ particles. Such a polydisperse system has been shown to consistently avoid crystallization, instead undergoing kinetic arrest into a `hard sphere glass' as the volume fraction is increased \cite{HSBerthier2016, HSOzawa2017}.  

Importantly, the hard sphere potential segments configuration space into two domains, a physical one with zero energy and a forbidden domain with one or more particle overlaps. To form a continuous energy landscape, we consider a `soft-sphere' extension to the HS model, with harmonic repulsion, see \hyperref[subsecSI:HS]{\textit{SI Appendix}} for details. This model thus resembles the SGM model, but has a different particle radius distribution and volume fraction. We then consider any configurations found with energy per particle less than a small tolerance ($U/N < U_{tol} = 10^{-16}$) to be physical HS configurations \cite{HSBerthier2016}. For such states with zero energy, we use a different order parameter to follow the progress of our algorithm across the landscape, corresponding to the mean coordination or contact number $\langle z \rangle$. Since pair contacts containing unconstrained or `rattler' particles can be trivially relaxed, $\langle z \rangle$ is a measure of the force bearing contacts. Its value is confirmed to be insensitive to the choice of $U_{tol}$, see \hyperref[subsecSI:HS]{\textit{SI Appendix}}, \Cref{figSI:HS}.

We perform MIMSE relaxation experiments varying the volume fraction initially over a small range, $0.66 < \phi < 0.67$. To provide initial ISes, we FIRE relax random configurations. Since this range is above the `jamming' or random close packed volume fraction for this model, $\phi_J^{HS} \simeq 0.65$, this initial quench step consistently yields ISes having finite energy, residing on the soft sphere portion of the landscape. Applied to these quenched states, we find MIMSE again rapidly descends the energy landscape, \Cref{fig:HSa}, when the bias radius is in a small range ($3 \lesssim \mathcal{U}_{\sigma} \lesssim 5$), indicating that there are canyons similar to those in the SGM landscape.

For the lower range of volume fractions we study, $\phi \lesssim 0.667$, MIMSE is able to consistently reach states with zero energy, corresponding to hard sphere configurations, \Cref{fig:HSa}. The first HS configurations found are isostatic (or nearly so), with $\langle z \rangle \approx 6$. Further application of biases briefly yields a mixture of HS and finite energy states, before reaching a part of the canyon where it samples a continuous series of HS states. Continuing MIMSE results in configurations with $\langle z \rangle$ dropping by roughly three orders of magnitude after $\sim 10^3$ biases, \Cref{fig:HSa}. Notably, when sampling the zero-energy domain, MIMSE finds a new, unbiased zero-energy state after every bias addition, and no finite energy ISes. The $g(r)$ of these configurations on the $\langle z \rangle$ plateau resembles the jammed case (for larger $r$ values) suggesting that the algorithm is sampling a region of the landscape corresponding to high pressures, adjacent to the jamming line \cite{HSBerthier2016, GlassEdiger2021}.

\begin{figure}[t]
\centering
\begin{captivy}{\includegraphics[scale=1]{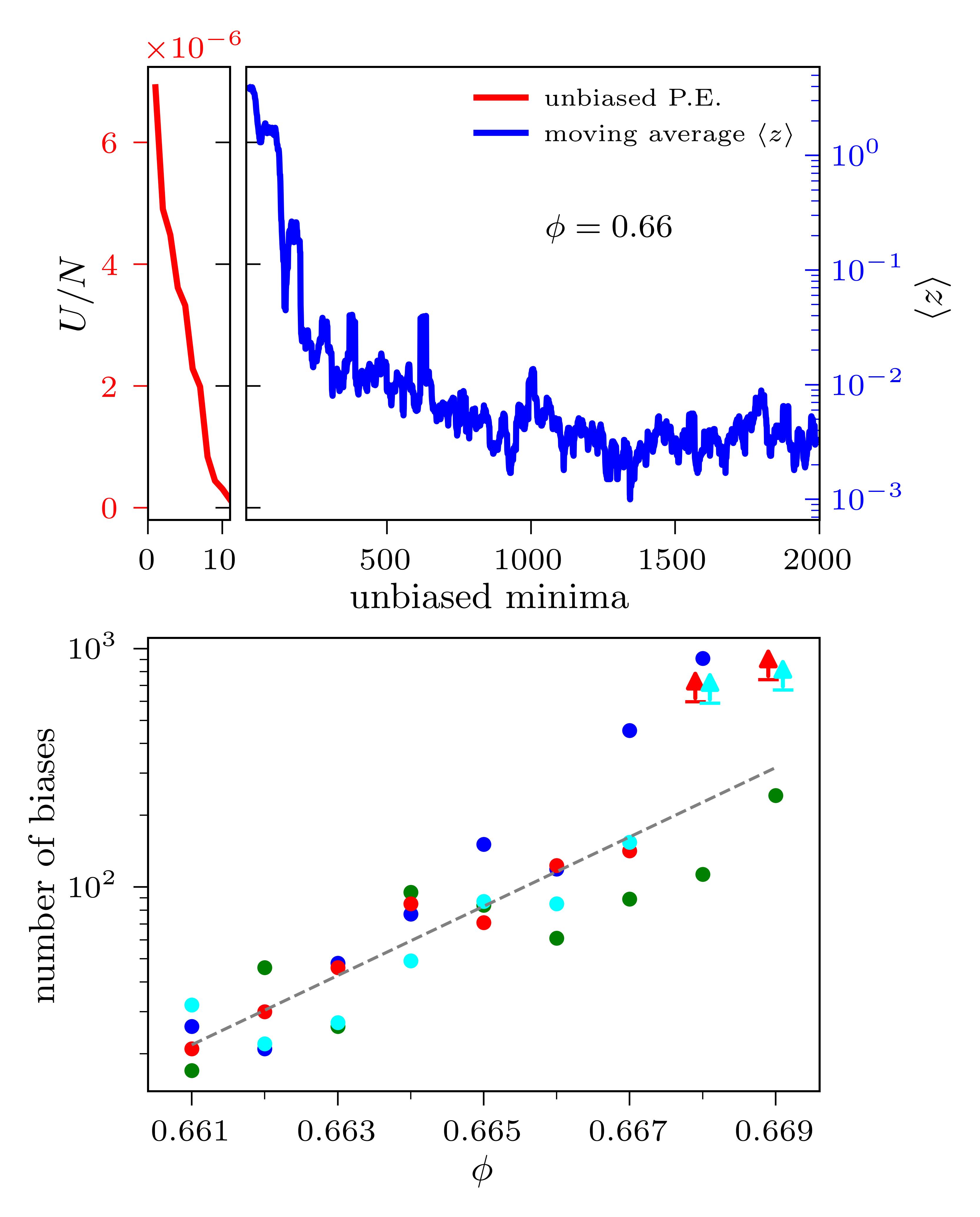}}
    \oversubcaption{0.05, 0.98}{}{fig:HSa}
    \oversubcaption{0.05, 0.48}{}{fig:HSb}
\end{captivy}
\caption{MIMSE descends both the soft sphere and HS portions of the HS glass landscape. (a) MIMSE efficiently descends down canyons, reducing the energy (left panel) of the soft-sphere system, ending in a nearly isostatic, `jammed' HS state. Further application of biases (right panel), shows that the average coordination number of these zero-energy configurations, $\langle z \rangle$, drops significantly, indicating structural relaxation, before reaching a steady state value.
(b) Running MIMSE on a range of different starting volume fractions shows us that MIMSE requires a larger number of biases to reach the HS portion of the landscape, becoming very computationally expensive for $\phi \gtrsim 0.667$. The dashed line is an exponential eye-guide. The arrow-headed lines denote lower-limit derived from incomplete, long simulations.}
\label{fig:HS}
\end{figure}

The behavior of the MIMSE algorithm while exploring such HS states provides insights about their distribution in configuration space, summarized in \Cref{fig:landscapeb}. Starting from an initial quenched configuration, MIMSE descends a finite energy canyon until it first reaches a portion of the canyon floor containing small `puddles' of zero energy, before reaching a region where nearly all ISes are covered by a slender, connected domain of HS configurations. It is as if the canyon floor were flooded by a `stream' or level set having zero energy. In this case, for every added bias, the canyon walls funnel all FIRE paths to points on the intersection of the edge of the bias and the edge of a zero energy domain, see \Cref{fig:landscapeb}.  FIRE then halts when the total energy per particle first drops below $U_{tol}$. The fact that the hyperspherical edge of every bias intersects a zero energy domain leads us to conclude that the domain is continuous (at least on the bias length scale used to search the configuration space).

As $\phi$ increases, MIMSE must work harder to reach HS states. Starting from a random `quenched' configuration, the number of biases required to reach the HS states (proportional to the contour distance travelled along the canyon floor) increases exponentially with $\phi$, \Cref{fig:HSb}. Notably, some initial configurations (and canyons) appear `deeper' than others, with a broad dispersion in the amount of effort to reach the HS states. Indeed, for $\phi \gtrsim 0.667$, MIMSE does not consistently reach HS states with a reasonable computational effort. It is as if the entire landscape has been raised and the `stream' drained away to a deeper, unreachably distant stretch of the canyon. In the opposite limit, the jamming volume fraction $\phi_J^{HS}$ corresponds to lowering the landscape so far as to flood the entire canyon, such that the HS domain covers even the highest, `quenched' ISes on the landscape. Overall, the finite energy portions of the canyons resemble that seen in the SGM case, \Cref{fig:landscapea}. Moreover, the number of biases between subsequent ISes on the soft-sphere portion of the landscape increases with $\phi$ and the number of FIRE steps required per bias increases by roughly ten-fold for $\phi \gtrsim 0.666$. This suggests a $\phi$-dependent change in the ruggedness of the landscape. 

Our results thus far finding canyons leading `deeper' into the landscape begs the question of how far (or deep in $\phi$) do the HS domain streams in these canyons go? Theoretically, the entropy of the equilibrium HS fluid is expected to nearly vanish (become sub-extensive) at a HS glass transition corresponding to $\phi_K^{eq}$. An extrapolation of simulation data in this system \cite{HSBerthier2016} suggests that this occurs at $\phi \approx 0.672$, only slightly higher than reached by MIMSE above. Thus it is natural to conjecture that the canyons we are exploring might cease to contain HS states (or `run dry') for some higher value of $\phi$. 

To reach deeper portions of the canyons, and find higher volume fraction HS states, we combined MIMSE with an adapted affine compression/relaxation scheme \cite{CompressionXu2005, JammingDesmond2009}, see \hyperref[subsecSI:HS]{\textit{SI Appendix}} for details. This method does consistently push the configuration further down the canyon, and reach higher $\phi$ states, again limited by computational effort. This approach typically yields HS states with volume fractions of $\phi \approx 0.670 \pm 0.001$. Remarkably however, two of $30$ runs reached dramatically higher volume fractions, $\phi = 0.681$ and $0.691$, the latter roughly $0.04$ above $\phi_J^{HS}$! Such dense configurations have previously been generated using swap Monte Carlo combined with compression methods \cite{HSOzawa2017}. Notably, unlike swap, the particle displacements used by our approach resemble physically allowed moves. Our findings thus suggest that a significant fraction of random `quenched' configurations (at least $2/30$) are connected to these ultra-dense states by physical trajectories.

To explore the landscape around the MIMSE-compression generated states, we used them as initial states for MIMSE runs, dilating them by various amounts. Without dilation, MIMSE quickly jumps out of the HS domain, sometimes finding a few adjacent HS states, and then only sampling soft-sphere ISes. This suggests that rather than a `stream' the HS domains are `puddles' smaller than or comparable in size to the bias radius. When the configuration is dilated by $\Delta \phi \approx 1-2 \times 10^{-3}$, however, MIMSE again finds long streams of HS states (and familiar canyon walls) that are easy to navigate (finding a new HS IS after every bias addition) adjacent to all the configurations including the one at $\phi = 0.681$, but not the densest $\phi = 0.691$ configuration. The landscape looks qualitatively different surrounding the latter state. This intriguingly suggests a change in the landscape for $0.68 < \phi < 0.69$, but clearly further study is required.

\section{\label{sec:KA}Kob-Andersen glass former}

Lastly, to determine if the canyons we find are also a feature of the energy landscapes of atomic glasses, we consider the Kob-Andersen (KA) model. This consists of a binary mixture having a total of $N = 256$ particles with $x_A = 0.8$ at a total number density of $\rho_{total} = 1.2$. To ensure force continuity, we employ a quadratically-smoothed, truncated form of the Lennard-Jones potential \cite{Stoddard1973}:
\begin{equation}
  V(\mathbf{r}_{ij})=\begin{cases}
    k\epsilon_{ij} \left({\left(\frac{\sigma_{ij}}{\lVert{\mathbf{r}_{ij}}\rVert}\right)}^{12} - {\left(\frac{\sigma_{ij}}{\lVert{\mathbf{r}_{ij}}\rVert}\right)}^6 \right) \\
    {} + \nu(\lVert{\mathbf{r}_{ij}}\rVert), & \text{if $\lVert \mathbf{r}_{ij} \rVert<\mathbf{r}_c$}\\
    0, & \text{otherwise},
  \end{cases}
\label{eq:KA-LJ potential}
\end{equation}
where $\epsilon_{AA} = 1.0$, $\epsilon_{BB} = 0.5$ and $\epsilon_{AB} = 1.5$; $\sigma_{AA} = 1.0$, $\sigma_{BB} = 0.88$ and $\sigma_{AB} = 0.8$, and $\nu(\lVert{\mathbf{r}_{ij}}\rVert)$ is a smoothing function (see \hyperref[subsecSI:KA]{\textit{SI Appendix}} for details). As before, we first choose random points in configuration space and FIRE relax them to their first energy minimum to form a quenched ensemble of ISes (\Cref{fig:KAa}, left panel).

When an optimal bias radius is used, we find that MIMSE efficiently descends to low energy portions of the KA landscape, \Cref{fig:KAa}. As in the SGM case (\Cref{fig:SGM}) the algorithm also proceeds to lower energy roughly logarithmically over the ISes sampled, despite the KA landscape's presumed differences from the earlier case. As before, the maximum bias radius is much smaller than the contour length of the configuration space path, consistent with a meandering canyon-like structure.  Descent, however, occurs in a much narrower window of bias radius, $1.0 \lesssim \mathcal{U}_\sigma \lesssim 1.5$. The KA system also requires far more biases for each new IS sampled. For the optimal bias parameters used, the distribution of the number of biases per new IS is heavy-tailed with a median of $\approx 9$ and $\langle n \rangle$ of $\approx 27$, versus $\langle n \rangle \approx 1.4$ in the SGM case. In addition to requiring more biases on average for each new IS, the algorithm must occasionally apply a large number of biases ($n>10^3$) to generate a new IS. Examination of the trajectories reveals the configuration path doubling back upon itself multiple times before crossing energy barriers. We interpret these differences from the SGM system as the landscape of the KA system being more rugged, requiring the filling of sub-basins (having a broad range of hypervolumes) with biases to cross the energy barriers necessary to find each new IS. We typically terminate the MIMSE run after a fixed amount of computational effort.

\begin{figure}[t]
  \centering
  \begin{captivy}{\includegraphics[scale=1]{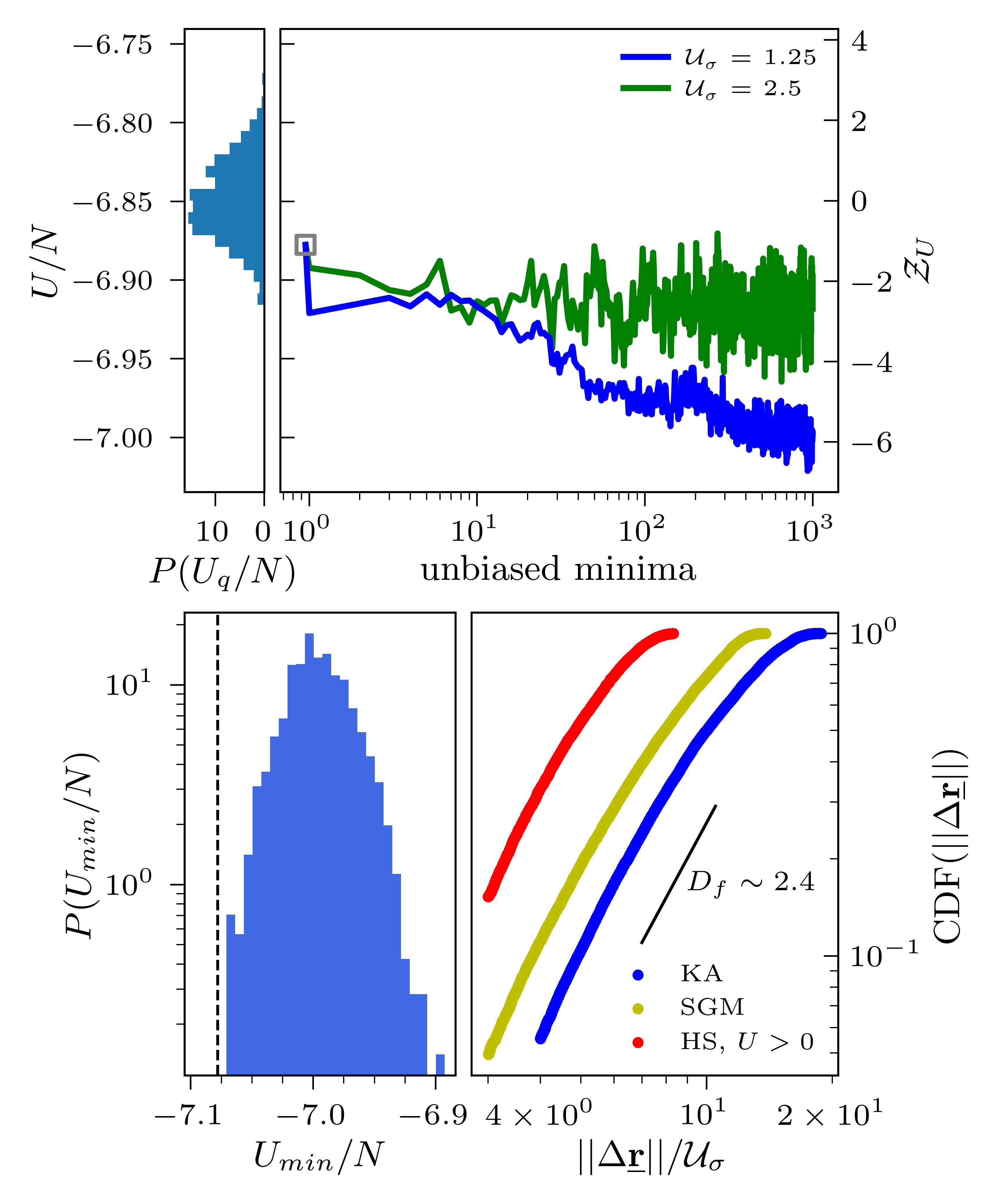}}
    \oversubcaption{0.05, 0.98}{}{fig:KAa}
    \oversubcaption{0.05, 0.48}{}{fig:KAb}
    \oversubcaption{0.95, 0.48}{}{fig:KAc}
    % \oversubtext{0.578, 0.6075}{}{\fontsize{6pt}{6pt}\selectfont
    % \cite{KauzmannAngell1999}}
    % \oversubtext{0.612, 0.637}{}{\fontsize{6pt}{6pt}\selectfont
    % \cite{KauzmannAngell1999}}
  \end{captivy}
  \caption{MIMSE descends the landscape of the Kob-Andersen glass former.
  (a) The left panel shows the probability distribution of energies for an ensemble of $1000$ `quenched' initial configurations, formed by FIRE relaxation of random configurations. On the right panel, we find that MIMSE successfully reaches low energy configurations for bias radii, $\mathcal{U}_\sigma \approx 1$. Both runs used $\mathcal{U}_0$ of $20$. 
  (b) Plotting a distribution of minimum energies from $\sim 1000$ runs shows that it closely approaches the Kauzmann limit (dashed line) \cite{KauzmannAngell1999}.
  (c) The cumulative density function (CDF) \cite{fractalGrassberger1983} of the pairwise separations of the ISes for all three systems show that they form low-dimensional clusters with very similar effective dimension: $D_f \approx 2.4$. Simulation parameters were $\mathcal{U}_0 = 3, \mathcal{U}_\sigma = 3$, (SGM); $\phi = 0.669$, $\mathcal{U}_0 = 0.85, \mathcal{U}_\sigma = 3.98$, (HS system with $U>0$); and $\mathcal{U}_0 = 20, \mathcal{U}_\sigma = 1.25$ (KA). The non-uniform sampling of the ISes was corrected as explained in \hyperref[secSI:fractalanalysis]{\textit{SI Appendix}}.}
  \label{fig:KA}
\end{figure}

The computational effort required to sample low energy states compares favorably to conventional methods. Earlier studies using simulated annealing of the KA model \cite{SimulatedAnnealinSastry1998} provide a relationship between IS energy and fictive equilibrium temperature. Characteristic temperatures such as the mode-coupling and Kauzmann temperatures can thus be mapped to their corresponding IS energies \cite{KauzmannAngell1999}. Benchmarking our algorithm against a Langevin dynamics implementation of simulated annealing finds a $\approx 100$-fold improvement in computational speed to reach configurations near the mode-coupling energy ($U/N \approx -7.00$).

Repeated MIMSE relaxations from a number of different starting configurations yields a roughly Gaussian distribution of minimum energies, \Cref{fig:KAb}.  Since these runs are performed for a fixed computational effort, this distribution is kinetically defined by our algorithm, and not a meaningful feature of the landscape itself. A structural analysis of these low energy configurations using common neighbor analysis \cite{aCNAStukowski2012, OVITOStukowski2009} confirms that they are amorphous. An earlier study using `basin-hopping' minimization \cite{BHWales1997} yielded a sample of similar energy values for amorphous structures in the same system \cite{GlassKAWales2001}, suggesting that MIMSE is acting like existing optimizers. The empirically predicted value of the Kauzmann energy, $U_K$ \cite{KauzmannAngell1999}, is about $\sim 3$ standard deviations below the mean of our distribution, $U_K/N \approx -7.08$,. Sampling several hundred meta-basins yields ISes very close to but not below the Kauzmann energy, consistent with this empirical prediction, see \Cref{fig:KAb}. For comparison with conventional methods, a lengthy Langevin dynamics simulation equilibrated at around the mode-coupling temperature \cite{KauzmannAngell1999}, yielded a much narrower distribution of inherent structure energies when quenched, $U/N \simeq -7.00 \pm 0.008$ and no states anywhere near the predicted Kauzmann limit.

The fact that the algorithm can fill sub-basins in the canyon floor implies that the low energy configurations occupy a low-dimensional subspace embedded in the higher dimensional configuration space. To estimate the effective dimensionality of the canyon floor, we compute the fractal dimension of the ISes we find, correcting for their non-uniform sampling using methods described in \hyperref[secSI:fractalanalysis]{\textit{SI Appendix}}. Specifically, we compute the correlation dimension \cite{fractalGrassberger1983} of a sub-ensemble of ISes; i.e., the scaling exponent of the cumulative distribution function of the $3N$-dimensional Euclidean distances between all pairs of ISes. If the ISes form a fractal, the correlation dimension would report the fractal dimension, $D_f$, via $CDF \sim \lVert{\Delta\underaccent{\bar}{\mathbf{r}}}\rVert^{D_f}$. The roughly power-law scaling seen, \Cref{fig:KAc}, is consistent with a fractal dimension, $D_f \approx 2.4$ \cite{fractalGrassberger1983}. The slight curvature of the plot suggests `multifractal' rather than true self-similar scaling. Unexpectedly, the observed multifractal geometry of the KA canyon floor appears to be very similar to that observed for canyon floors in the SGM and HS systems, \Cref{fig:KAc}. Further analysis of the SGM system confirms a cross-over to a slightly higher effective dimension at shorter lengthscales, see \hyperref[secSI:fractalanalysis]{\textit{SI Appendix}}. 

\section{\label{sec:Discussion}Discussion}

We have found that our metadynamics algorithm provides interesting insights into the large-scale structure of three glassy energy landscapes, in particular the canyon-like subspaces that both contain the glassy configurations and provide direct routes leading to the landscapes' low energy states. While the `floors' of these canyons contain dense clusters of inherent structures and resemble the well-known rugged and barrier-filled landscapes of glasses, our work reveals that these glassy domains are surrounded by large, high-dimensional `canyon walls' effectively devoid of minima and barriers, and the canyons contain energy gradients that lead directly to lower energies in the landscape. Such simply connected routes between glassy states and the lowest energy states pose a conundrum---how is it that simulated annealing struggles to find very low energy states? The answer is presumably a free energy barrier, as suggested by our finding that the canyons in the SGM landscape become narrower at lower energies in the landscape. In the HS case, we are tempted to associate the `streams' of hard sphere states we found with the results of Ref. \cite{BasinAshwin2012} for small packings of hard spheres; they found most hard sphere states comprised long, narrow `threads' that terminate in high-density `cores' (which themselves contain very few states). Our finding of narrow `streams' in the landscape up to at least $\phi = 0.681$, however, seems incompatible with the prediction that the HS entropy becomes sub-extensive at a lower value. In the Kob-Andersen glass former, the algorithm is able to consistently reach low-energy portions of the canyon, including energies approaching the predicted Kauzmann limit. This finding confirms that by virtue of its ability to follow smooth canyon walls to lower energy, MIMSE can act like a global optimizer similar to basin hopping \cite{BHWales1997}. Finally, the presence of similar canyon-like, low-dimensional subspaces with very similar multifractal geometry in these seemingly different systems might explain the range of qualitatively similar physical phenomenon and kinetics in different glassy systems \cite{IdealGlassRoyall2018, GlassesDebenedetti2001, GlassEdiger2021}.

Future work will apply MIMSE to other glassy landscapes, including bonded model systems. While informative, MIMSE does not obey detailed balance, nor does it return samples corresponding to any canonical ensemble. We anticipate that future work will fruitfully hybridize MIMSE with other methods, such as swap \cite{HSOzawa2017, KAParmar2020} or ghost-particle \cite{GMCRoyall2017} Monte Carlo, and parallel tempering \cite{ParTempHansmann1997} to enable barrier crossing while obtaining canonical sampling. Recently the loss landscape of deep neural networks has been shown to have similar features \cite{NNChen2022} to that of a soft glassy matter system, perhaps MIMSE will enable useful exploration of such landscapes as well.

\begin{acknowledgments}
% \section*{Acknowledgments}
% \vspace{2mm}
We are grateful for useful conversations with Andrea Liu and Talid Sinno. This work was supported by NSF-DMR 1609525 and 1720530 and computational resources provided by XSEDE through TG-DMR150034.
\end{acknowledgments}

\appendix

\section{\label{secApp:methods}Details of the Algorithm}
% \subsection*{Implementation Details}
A detailed description of the algorithmic procedure is provided below.
\begin{enumerate}[noitemsep]
  \item A random energetic minima sampled via FIRE from a random configuration (quenched ensemble) is selected as the starting point.
  \item A bias potential is added on to the system centered around the selected minimum in 3N dimensional configuration space. (Note: to ensure the bias potential is center of mass preserving during evaluation of the bias, we subtract the mass weighted sum of forces on all particles.)
  \item  The system is given a small, positional displacement in a random direction in the configuration space, scaled such that at least one particle exceeds the system force tolerance, $F_{tol}$. This displacement is adjusted to preserve the center of mass location.  The system is then quenched on the biased energy landscape (${U}_\text{tot} = {U} + {U}_{\text{i, bias}}$, where $i$ signifies the bias number)
  \item Biases are added in succession until a new unbiased minimum is reached. This is verified by ensuring that ${U}_{\text{total, bias}}/ N < U_{tol}$.
  \item The above process is repeated until a desired amount of the energy landscape is sampled.
\end{enumerate}

To undertake the above process with utmost efficiency and accuracy for our systems of interest with a highly rugged energy landscapes, we use a finite-ranged bias potential viz. a symmetric quartic function, as described in \Cref{eq:bias potential}. Parameters $\mathcal{U}_\sigma$ and $\mathcal{U}_0$ represent the $3N$ dimensional Euclidean extent and energetic height of the bias respectively. Throughout our simulation, the center of mass of the system remains at rest, due to the bias force exerting no force on the center of mass. This is enforced by subtracting the mass weighted total force from individual bias force calculations. Despite this correction, our bias remains isotropic in the $3N-1$ angular dimensions.

We select a pair of parameters that efficiently samples lower energy states effectively and does so with a small number of biases per every new (unbiased) minima. We use a divide and conquer-like search procedure to determine a suitable pair of parameters for our algorithm. Our initial guess for both parameters are based on the potential energy of interaction. For example, in the SGM system, we search for an optimal $\mathcal{U}_0$ between $\sim \epsilon /2$ and $\sim N^2 \epsilon/ 2$. Meanwhile, a search between and around $\sim \langle R \rangle$ and $\sim \sqrt{N}\langle R \rangle$. The selection of parameters is further verified by studying the gradient and clusters formed by trajectories emerging from a starting IS using different $\mathcal{U}_\sigma$ values (see \Cref{fig:SGMcluster}).

To keep track of the biases, we use a modified extended $3N$-dimensional neighbor list that keeps track of all biases in high dimensional space around the system configuration within a cut-off distance $\mathcal{U}_{c1}$. Further, we have a long range secondary cut-off length $\mathcal{U}_{c2}$ which defines our neighbor search boundary; biases are retired when the system is farther than $\mathcal{U}_{c2}$ from a particular bias. Thus, in effect we maintain neighbors (biases) within $\mathcal{U}_\sigma + \mathcal{U}_{c1}$, and update the neighbor list by searching for biases within $\mathcal{U}_{c2}$ from the system position. The update is done when the system moves an Euclidean displacement more than $\mathcal{U}_{c1}$, whose value is chosen for computational efficiency. Parameter $\mathcal{U}_{c2}$ is chosen so that the ultimate trajectory of the system is not affected. It may be noted that both of these parameters are system and bias size dependent, and need to be tuned for each system to balance the efficiency of the method with not retiring biases prematurely.

Our metadynamics-based algorithm differs from that developed by Yip in three key ways: (i) We use a smooth, truncated bias potential \Cref{eq:bias potential} instead of the traditional Gaussian bias, that enables the use of a high dimensional neighbor list. This allows us to significantly reduce the computational overhead associated with computing the total bias potential, and also facilitates the retirement of biases beyond a certain distance. (ii) Further, we make sure that the center of mass of the system remains at rest throughout the simulation. We enforce this on the bias potential and the random displacement after bias addition, which appears to lead to significantly more efficient descent down the landscape. (iii) Last, we use an optimized minimzer - FIRE \cite{StrucRelaxBitzek2006}, that helps boost the computational efficiency while exploring such glassy high-dimensional landscapes.

These modifications and additions allow us to overcome the challenges posed by our high dimensional system. It must be noted that each system responds differently to the biases in a way that depends on the characteristic energy and length scales of that particular system.   

\section{\label{secSI:ModelDetails}Model Details}

\begin{figure}[t]
\centering
\begin{captivy}{\includegraphics[scale=1]{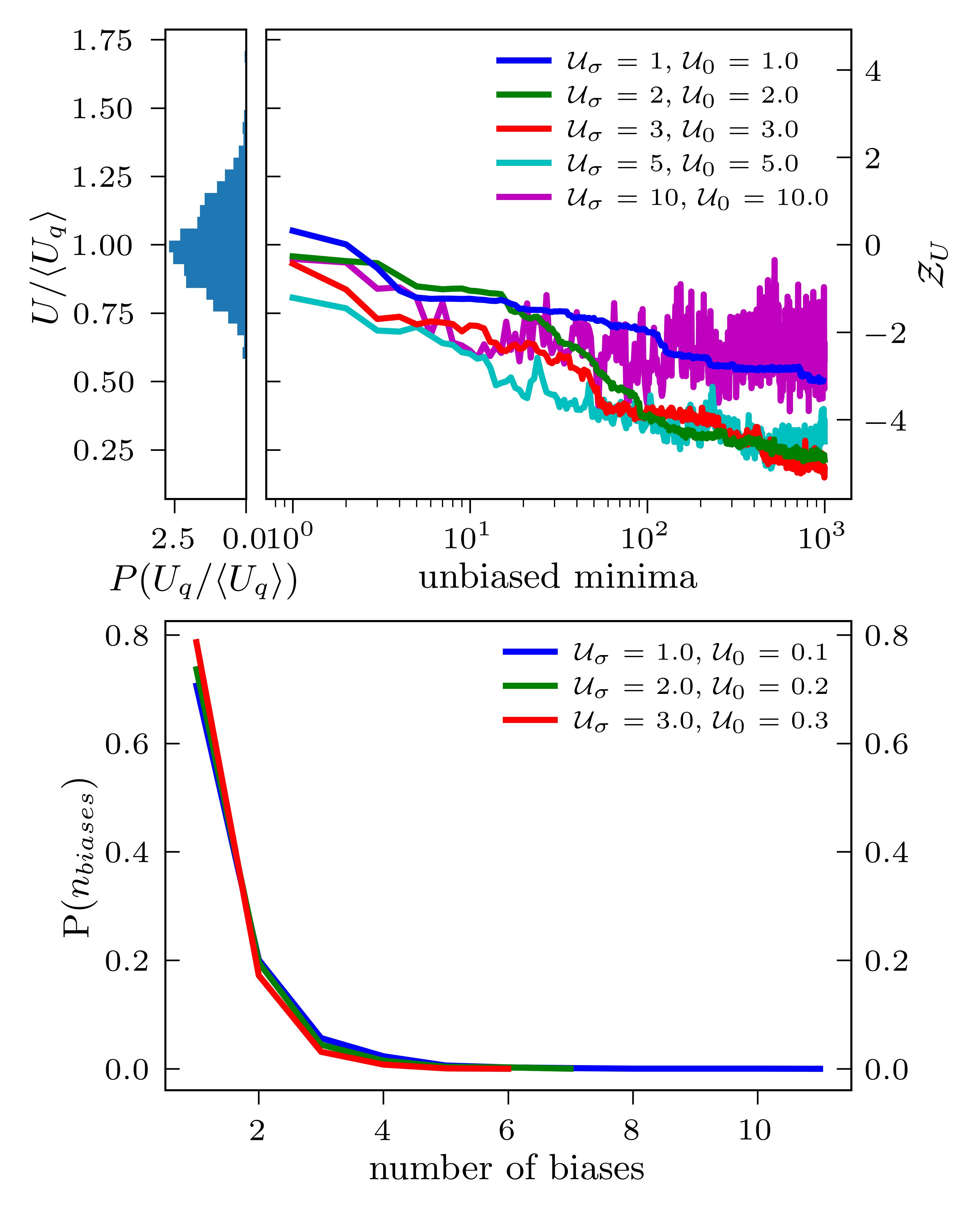}}
    \oversubcaption{0.05, 0.98}{}{figSI:SGMa}
    \oversubcaption{0.05, 0.48}{}{figSI:SGMb}
\end{captivy}
\caption{\label{figSI:SGM} (a) Energy descent versus bias radius $\mathcal{U}_{\sigma}$ confirms that SGM landscape canyons are tapered. The optimal choice of $\mathcal{U}_{\sigma} =3$ descends to significantly lower energy states. Smaller biases also descend, but more slowly. Larger biases descend rapidly at first, but plateau at an energy when they have become too big for the narrowing canyon. (b) The probability distribution of the number of biases that must be applied between two consecutive ISes appears insensitive to $\mathcal{U}_{\sigma}$, suggesting a self-similar rather than uniform distribution of ISes along the MIMSE descent path.}
\end{figure}

\subsection{\label{subsecSI:SGM}SGM Model}
The Soft Glassy material system is a set of $N = 342$ particles interacting via
\begin{equation}
 V(\mathbf{r}_{ij})=\begin{cases}
    \frac{\epsilon}{\alpha} {\left(1 - \frac{\lVert{\mathbf{r}_{ij}}\rVert}{R_i + R_j} \right)}^{\alpha}, & \text{if $\lVert \mathbf{r}_{ij} \rVert<R_i + R_j$}\\
    0, & \text{otherwise},
 \end{cases}
\label{eq:soft-sphere potential}
\end{equation}
with $\alpha=2$. The particles have a radii distribution resembling a Weibull distribution, $\textbf{P}(R) \sim (k/ \lambda) (x/ \lambda)^{k-1}$, where $k = 1.66, \lambda = 0.78$. The exact distribution snapshot is taken from a model foam simulation \cite{SGMHwang2016}. To ensure meaningful descent, we further tweak our algorithm so that the biases only act on non-rattler particles (those having a coordination $z \geq d+1=4$), which significantly enhances descent. We use tolerance values of $U_{tol} \sim 10^{-12}$ and $F_{tol} \sim 10^{-7}$ for simulations for this system. As noted in the main text, for slightly too large values of $\mathcal{U}_\sigma$ the algorithm gets stuck, plateauing at intermediate energy, \Cref{figSI:SGMa}. This confirms the interpretation that the canyon is tapered as it descends to lower energies.

As the algorithm proceeds we can track the number of new biases $n$ required to produce each unbiased IS. Unexpectedly, both $\langle n \rangle \sim 1.4$ and the distribution $P(n)$ are similar and insensitive to the bias radius $\mathcal{U}_\sigma$, \Cref{figSI:SGMb}. If the ISes were distributed uniformly along the algorithm's path, larger biases would be more likely to find biased minima, and $\langle n \rangle$ would increase with $\mathcal{U}_\sigma$. This insensitivity can be naturally explained by structure of the IS cluster being self-similar on these length scales.

\subsection*{\label{subsecSI:ClusterAnalysis}Cluster Analysis}
We use a single-linkage clustering analysis \cite{SLCAGower1969} to analyze the connectivity of the different $\hat{u}$ obtained through the $1100$ ($>3N$) different trajectories as a function of $\mathcal{U}_\sigma$. We find that using an angle of $\theta_0 \sim 1.31 \text{rad}$ as the linkage threshold suffices to connect the $\hat{u}$ in the best case. We see that for other angles greater and smaller than $\theta_0$, the ISes form a single or large number of clusters. That is, we find that at $\theta_0$, the system shows a clear trend for the number of clusters detected (see \Cref{fig:SGMclusterb}). We also ignore any outliers detected as small or singleton clusters in the analysis using the following criteria: $n_{cluster} \leq 0.25 \times N_{points}/N_{cluster}$ \cite{OutlierSV2015}. It may also be noted that the bias potential acts on all particles, unlike the descent case explained earlier in the SGM section.

\subsection{\label{subsecSI:HS}HS glass}
To study this system, we use a well-studied \cite{HSBerthier2016, HSOzawa2017, EntropyBerthier2017} model of hard-sphere (HS) glasses, having a size distribution given by: $\textbf{P}(R) \sim R^{-3}, R \in [R_{\text{min}}, R_{\text{max}}], R_{\text{min}}/R_{\text{max}} = 0.4492$, and $N=1000$ particles; and $R_{\text{max}}$ is taken as the reference $=1$.

\begin{figure}[ht]
  \centering
    \includegraphics[scale=1]{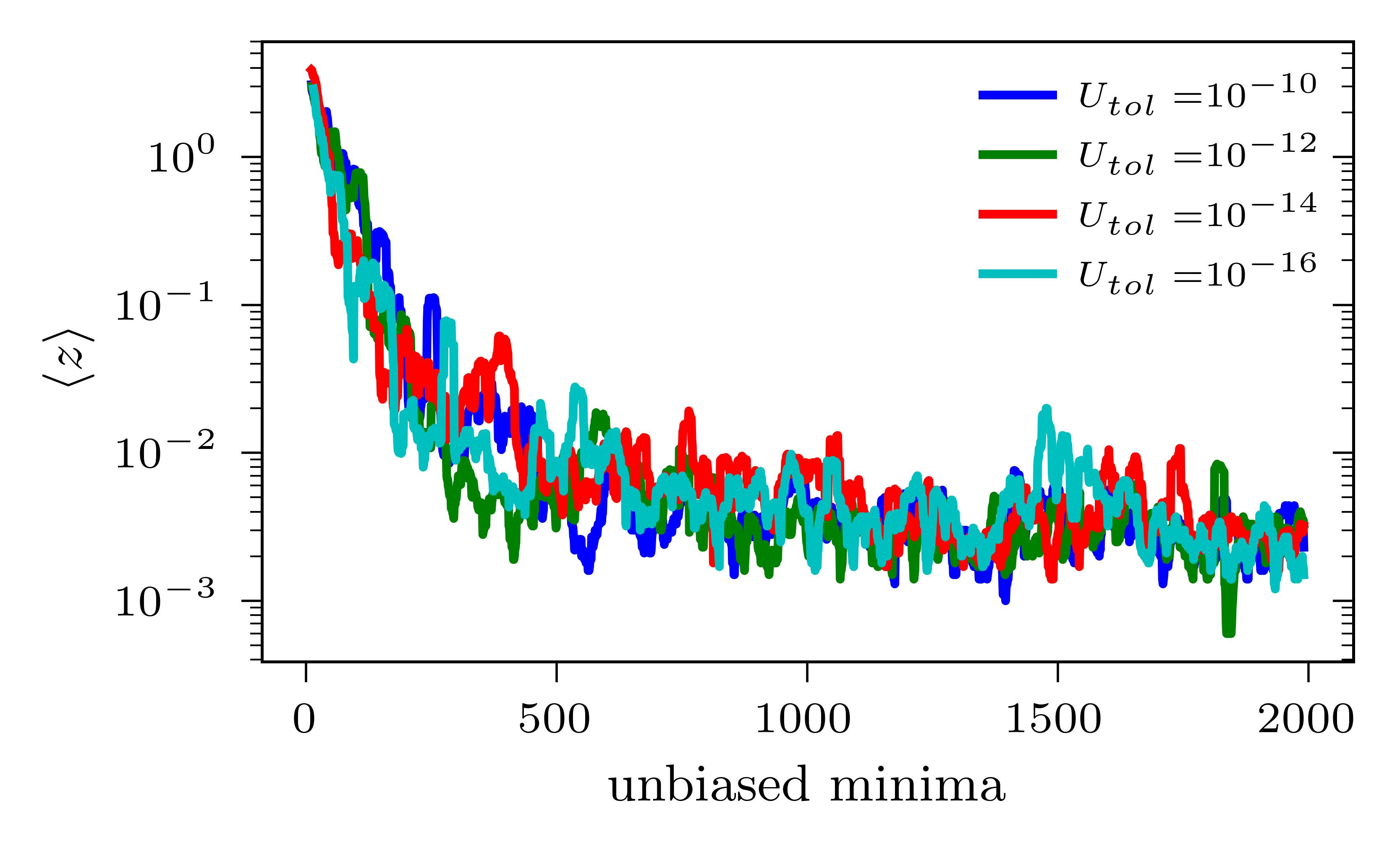}
\caption{\label{figSI:HS} We use contact coordination, $\langle z \rangle$, as an order parameter for hard sphere configurations (see \Cref{fig:HSa}).  We verified that this order parameter is physically sensible by finding that its value is insensitive to the choice of energy tolerance, $U_{tol}$.}
\end{figure}

To compare our results with earlier work \cite{HSBerthier2016}, we use a similar tolerance value of $U_{tol} = 10^{-16}$. We further use a $F_{tol} \sim 10^{-7}$ to determine positional displacements in MIMSE. MIMSE works efficiently on the HS portion and the soft sphere extension of the landscape. All results (\Cref{fig:HS}) discussed and shown are normalized with respect to the $\phi_{ref}=0.66$ system. We use $\mathcal{U}_{\sigma,ref} = 4$ and $\mathcal{U}_{0,ref} = 0.1$ as reference bias dimensions and scale them with changes in $\phi$. The scaling relation used to determine bias parameters at $\phi$ are: $\mathcal{U}_0 = \frac{\mathcal{U}_0,ref}{U_{IS, ref}} U_{IS}$, and $\mathcal{U}_{\sigma} = \frac{\mathcal{U}_{\sigma}}{\phi^{(1/3)}} \phi_{ref}^{(1/3)}$; where $U_{IS}$ represents the IS energy obtained through random quenching at $\phi$.

In \Cref{fig:HSa}(right panel), we use the average coordination number $\langle z \rangle$ as an order parameter to study the HS state exploration facilitated by MIMSE. We find that the effective $\langle z \rangle$ values remain independent of $U_{tol}$, see \Cref{figSI:HS}, suggesting that it is a reliable feature of the configuration and not the algorithm.

To create HS states at high volume fraction we combine MIMSE with an affine compression and relaxation scheme by Weeks \cite{JammingDesmond2009}.  Specifically, we run the compression scheme with FIRE as a minimizer until the system jams and is unable to reach higher volume fractions. We then run MIMSE on that jammed configuration, reducing its contact coordination significantly, as in \Cref{fig:HSa}.  Rerunning the compression scheme then allows a still higher volume fraction to be reached before the system again jams. This alternating compression and unjamming (using MIMSE) process, which we call MIMSE-compression, was repeated to attain HS states up to $\phi \approx 0.670$. Two of 30 MIMSE-compression runs produced surprisingly dense configurations: $\phi \simeq 0.691$ and $\phi \simeq 0.681$. 

To explore the energy landscape around the densest state, we dilate the configuration to a range of volume fractions and run MIMSE on the corresponding energy landscape. Initially, the algorithm finds a number of zero-energy states at all volume fractions, until it covers a small domain around the initial configuration. After that point, however, in this region of configuration space the algorithm finds only biased minima, which is distinctly different than the landscape explored at lower volume fractions. In principle, the preponderance of biased minima could represent two distinct cases on the landscape. The first possibility is that of a `rugged' landscape, where the low-force edge of the bias is filled with many ISes, which trap the configuration before it can leave the biased domain. The other possibility is that of a `tortuous' landscape, where the path taken by FIRE would be non-monotonic in radial distance from the bias center, requiring climbing in energy, and resulting in minima stabilized by the bias force. The two cases can be distinguished by the energy change upon relaxing the biased minimum configuration with the biases removed, which is either a very small change (rugged case) or a large one (tortuous case).  Adjacent to the high density state, the landscape appears highly `tortuous' below $\sim 0.67$, and `rugged' above that value. Lastly, MIMSE in this portion of the landscape only finds HS states for $\phi \lesssim 0.67$, and then only by performing FIRE relaxations from biased minima with all biases removed. While presumably a zero energy path still exists (i.e. the one traversed by compression to get to the dense state), MIMSE's inability to find it suggests a mismatch between the path and the bias parameters used. Further study is needed.

\subsection{\label{subsecSI:KA}KA glass former}
The Kob-Andersen (KA) model glass is a binary system with a total of $N = 256$ particles with number fraction $x_A = 0.8$ at a number density of $\rho_{total} = 1.2$ \cite{SimulatedAnnealinSastry1998}. To ensure force continuity, we employ a quadratically-smoothed, truncated form of the Lennard-Jones potential \cite{Stoddard1973}. We use a tolerance value of $U_{tol} \sim 10^{-16}$ and $F_{tol} \sim 10^{-7}$ to define IS minima in this model.

\subsection*{\label{subsecSI:SimAnnealing}Simulated Annealing}\textit{}: To compare MIMSE's ability to explore and generate low-energy states, we employ Langevin Dynamics based simulated annealing. The system is first equilibrated at $k_BT = 1.0-1.5$, and then systematically cooled using a linear cooling ramp. The thermal configurations are periodically quenched using FIRE to obtain corresponding ISes. Our results reproduce those in a previous study \cite{SimulatedAnnealinSastry1998}. Comparing the computational cost to generate low energy ISes using the two different methods shows $\approx 100$ fold speed-up when using MIMSE at final energy values of $U/N \approx -7.00$.

\begin{figure}[ht]
\centering
\begin{captivy}{\includegraphics[scale=1]{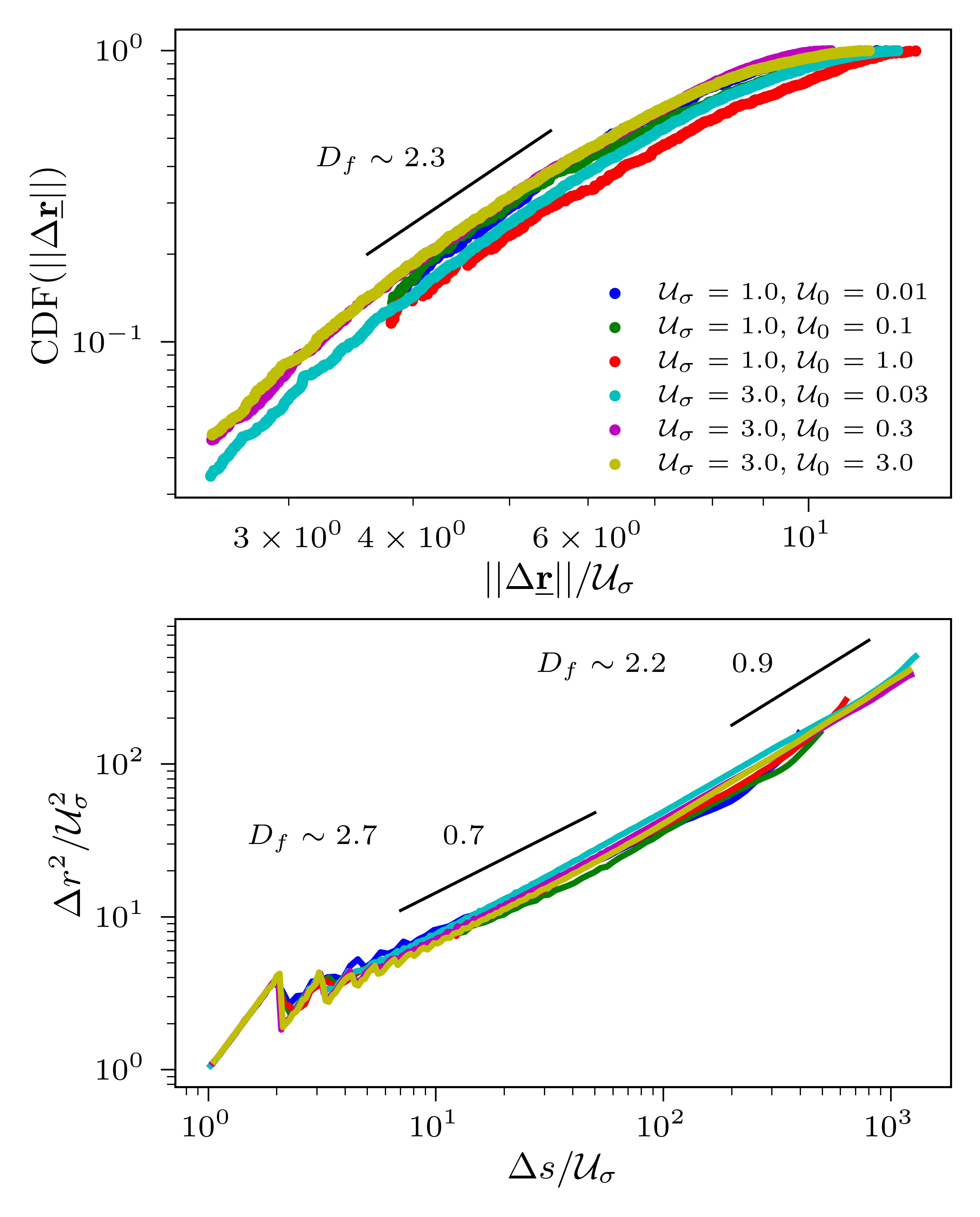}}
    \oversubcaption{0.05, 0.98}{}{figSI:SGMfractala}
    \oversubcaption{0.05, 0.48}{}{figSI:SGMfractalb}
\end{captivy}
\caption{\label{figSI:SGMfractal} Fractal scaling analysis of the IS ensemble sampled by MIMSE on the SGM landscape: (a) As explained in the text, the correlation dimension obtained on the over-sampling corrected ensemble of ISes gives a fractal dimension $\sim 2.3$ for the set of ISes, when averaged over $3$ such subsets, indicating the low dimensionality of the canyons floors explored by MIMSE. The subensembles were made using $l_{max}/ \mathcal{U}_\sigma$ values of $2.7$ and $1.9$ for $\mathcal{U}_\sigma$ values of $1$ and $3$, respectively. (b) Calculating the fractal scaling of the path corroborates the findings in (a), departures from a single power-law indicate a multifractal scaling with a range of fractal dimensions $2.3 - 2.7$.}
\end{figure}

\section{\label{secSI:fractalanalysis}Landscape fractal analysis}

The low energy portions of the potential energy landscape seem to occupy a low-dimensional subspace embedded in the high-dimensional configuration space. To estimate the dimensionality of this subspace for different landscapes and in different energy regions, we compute the effective fractal dimension of the ensemble of ISes sampled by MIMSE. The fractal dimension of a set of points (and their bounding subspace) can in principle be computing using the correlation dimension \cite{fractalGrassberger1983}; that is, the scaling exponent of the cumulative distribution function of the $3N$-dimensional Euclidean distances between all pairs of points in the ensemble. If the points were densely and uniformly distributed in the subspace, this process would be straightforward, but in general, sampling from MIMSE (or canonical sampling methods) is both non-uniform and has a finite number of samples.

To obtain a robust estimate of the fractal dimension of a volume bounding a set of points, we must determine (1) the minimum lengthscale over which the volume is well sampled and (2) ensure that we have not significantly oversampled small regions, which could undesirably bias the estimate for the correlation dimension at small lengthscales. For the first point, we form a single-linkage hierarchical cluster of the set points by creating a hierarchical dendrogram. We then assume that the maximum Euclidean distance ($l_{max}$) on the dendrogram (the longest edge required to just form a single cluster) is the smallest lengthscale over which we have reliable sampling. For the second point, we `declutter' the ensemble of points such that no two points are closer together than $l_{max}$. This is accomplished by randomly marking a point and discarding any other points that are closer to it than $l_{max}$, and then repeating the process with random unmarked points until all points have been marked or discarded. We then calculate the correlation dimension of the marked subsample. Typical results presented are averaged over $3$ random subensembles formed from a single ensemble of ISes. Applying this analysis to the ISes in the descending canyon of the three systems - SGM at $\mathcal{U}_0 = 3, \mathcal{U}_\sigma = 3$, Soft-sphere portions of the HS landscape at $\phi = 0.669$ with scaled $\mathcal{U}_0 = 0.1, \mathcal{U}_\sigma = 4$ (see ) and KA glass former at $\mathcal{U}_0 = 20, \mathcal{U}_\sigma = 1.25$ show that they form low-dimensional clusters with very similar fractal dimensions of $D_f \approx 2.4$. The ensemble of ISes was decluttered using $l_{max}/\mathcal{U}_\sigma$ values of $2.8$, $2.7$ and $2.5$ units for the SGM, HS and KA systems, described above, respectively.

We also analyzed the fractal geometry of the configuration space path sampled by MIMSE, using methods from an earlier study \cite{SGMHwang2016}. For each pair of ISes on the path both their Euclidean distance, $\Delta r^2$, and the contour length $\Delta s$ of the path between them are first computed, then pooled together, partitioned in $\Delta s$ and averaged. This yields the scaling of Euclidean distance with contour length, analogous to a mean-squared displacement in lower-dimensions. For a fractal curve, these quantities obey a power-law relationship, $\Delta r^2 \sim \Delta s^{(2/D_f)}$, where $D_f$ is the fractal dimension. This approach is reliable when the points are almost uniformly spaced and are ordered into a sequence; when MIMSE is nearly monotonically descending a canyon, this analysis gives results that are consistent with the method described above. However, when energy gradients are lower and/or the landscape is more complex/ rugged such that the algorithm doubles back on itself, this method gives unreliable results due to oversampling small lengthscales.

\Cref{figSI:SGMfractalb} shows the results for the configuration space path on the SGM landscape for different bias parameters, which resembles a crossover between two power-laws, indicating only rough self-similarity. At the longest scales, the configuration path scaling suggests $D_f^{MSD} \approx 2.3$, consistent with the previous method. This value is slightly higher than that for a random walk $D_f = 2$, suggesting that the canyon MIMSE is following is slightly more confined than a random walk. This analysis also allows scaling to be studied at smaller lengthscales; which appear to have a slightly higher effective fractal dimension, $D_f^{MSD} \approx 2.7$. It may be noted that these canyon fractal dimensions are very different than ones found ($D_f \sim 1.37$) for the same SGM system in the higher energy portions of the landscape \cite{SGMHwang2016}.

The small scale tortuosity of the configuration path explains the power-law dependence of the descent rate on $\mathcal{U}_\sigma$, \Cref{fig:SGMb} (inset). If MIMSE essentially follows the same path with different-sized biases, more smaller biases would be required to `cover' a tortuous path. This is analogous to another method for computing fractal dimension, which consists of counting the number of boxes required to cover a curve as a function of the box size. We suppose that the scaling exponent seen in \Cref{fig:SGMb}(inset) effectively provides such a `box-counting' measure of the path fractal dimension, $D_f^{bias} \approx 2.8$ which is in fact consistent with $D_f^{MSD} \approx 2.7$ at that length-scale.
% 
% \clearpage
% \nocite{*}

% \balance
% \typeout{}
\bibliography{arXiv/references}% Produces the bibliography via BibTeX.

\end{document}